\documentclass{sf2a-conf2016}
\usepackage{graphicx}
\usepackage{hyperref}
\usepackage[]{natbib}  
\usepackage{epstopdf}

\def\BibTeX{{\rm B\kern-.05em{\sc i\kern-.025em b}\kern-.08em
    T\kern-.1667em\lower.7ex\hbox{E}\kern-.125emX}}
\bibpunct{(}{)}{;}{a}{}{,}  

\begin{document}

\TitreGlobal{SF2A 2016}


\title{Understanding Active Galactic Nuclei using near-infrared high angular resolution polarimetry II: Preliminary results}

\runningtitle{AGN at high angular resolution: M\lowercase{ont}AGN - STOKES}

\author{F. Marin}\address{Observatoire Astronomique de Strasbourg, Universit\'e de Strasbourg, CNRS, UMR 7550, 11 rue de l'Universit\'e, 67000 Strasbourg, France}

\author{L. Grosset}\address{LESIA, Observatoire de Paris, PSL Research University, CNRS, Sorbonne Universit\'es, UPMC Univ. Paris 06, Univ. Paris Diderot, Sorbonne Paris Cit\'e}
\author{R. Goosmann$^1$}
\author{D. Gratadour$^2$}
\author{D. Rouan$^2$}
\author{Y. Cl\'enet$^2$}
\author{D. Pelat}\address{LUTH, Observatoire de Paris, CNRS, Université Paris Diderot, 92190 Meudon, France}
\author{P. Andrea Rojas Lobos$^1$}

\setcounter{page}{237}


\maketitle


\begin{abstract}
In this second research note of a series of two, we present the first near-infrared results we obtained 
when modeling Active Galactic Nuclei (AGN). Our first proceedings  showed the comparison between the 
MontAGN and STOKES Monte Carlo codes. Now we use our radiative transfer codes to simulate the 
polarization maps of a prototypical, NGC~1068-like, type-2 radio-quiet AGN. We produced high angular resolution 
infrared (1~$\mu$m) polarization images to be compared with recent observations in this wavelength range. 
Our preliminary results already show a good agreement between the models and observations but cannot account 
for the peculiar linear polarization angle of the torus such as observed. \citet{Gratadour2015} found a 
polarization position angle being perpendicular to the bipolar outflows axis. Further work is needed 
to improve the models by adding physical phenomena such as dichroism and clumpiness. 
\end{abstract}

\begin{keywords}
galaxies: active, galaxies: Seyfert, radiative transfer, techniques: polarimetric, techniques: high angular resolution
\end{keywords}


\section{Introduction}
Understanding the morphology, composition and history of each AGN component is a non-trivial goal that requires a strong 
synergy between all observational techniques. The role of polarimetric observations was highlighted in the 80s thanks to the 
discovery of broad Balmer lines and Fe~{\sc ii} emission sharing a very similar polarization degree and position angle with 
the continuum polarization in NGC~1068, a type-2 AGN \citep{Antonucci1985}. The resemblance of the polarized flux spectrum with 
respect to the flux spectrum of typical Seyfert-1s lead to the idea that Seyfert galaxies are all the same, at zeroth-order 
magnitude \citep{Antonucci1993}. Observational difference would arise from a different orientation of the nuclei between pole-on and 
edge-on objects; this is due to the presence of an obscuring dusty region situated along the equatorial plane of the AGN that will 
block the direct radiation from the central engine for observers looking through the optically thick circumnuclear material.
This is the concept of the so-called ``dusty torus'', first conceived by \citet{Rowan1977} and later confirmed by \citep{Antonucci1985}.

Since then, a direct confirmation for the presence and structure of this dusty torus was an important objective for the AGN 
community. The closest evidence for a dusty torus around the central core of NGC~1068 was first obtained by \citet{Jaffe2004} 
and \citet{Wittkowski2004}, using  mid-infrared (MIR) and near-infrared (NIR) interferometric instruments coupled to the European 
Southern Observatory's (ESO's) Very Large Telescope interferometer (VLTI). \citet{Jaffe2004} were able to spatially resolve the 
MIR emission from the dusty structure and revealed that 320~K dust grains are confined in a 2.1~$\times$~3.4~pc region, surrounding 
a smaller hot structure. The NIR data obtained by \citet{Wittkowski2004} confirm the presence of this region, with the NIR fluxes 
arising from scales smaller than 0.4~pc. Since then, long-baseline interferometry became a tool used to explore the innermost
AGN dusty structure extensively at high angular resolution (typically of the order of milli-arcsec, see e.g., 
\citealt{Kishimoto2009,Kishimoto2011}).

Coupling adaptive-optics-assisted polarimetry and high angular resolution observations in the infrared band, \citet{Gratadour2015} 
exploited the best of the two aforementioned techniques to obtain strong evidence for an extended nuclear torus at the center of 
NGC~1068. Similarly to previous optical \citep{Capetti1995} and infrared \citep{Packham1997,Lumsden1997} polarimetric observations, 
\citet{Gratadour2015} revealed an hourglass-shaped biconical structure whose polarization vectors point towards the hidden 
nucleus. By subtracting a purely centro-symmetric component from the map of polarization angles, an elongated (20~$\times$~50~pc) 
region appeared at the assumed location of the dusty torus. If the signal traces the exact torus extension, high angular resolution 
polarization observations would become a very powerful tool to study the inner core of AGN.

In this lecture note, the second of the series, we will show the preliminary results obtained by running Monte Carlo radiative 
transfer codes for a NGC~1068-like AGN. Our ultimate goal is to reproduce the existing UV-to-infrared polarimetric observations 
using a single coherent model in order to constrain the true three-dimensional morphology of the hard-to-resolve components of 
close-by AGN.

\section{Building an NGC~1068 prototype}
Our primary model is powered by a central, isotropic, point-like source emitting an unpolarized spectrum with a power-law spectral 
energy distribution $F_{\rm *}~\propto~\nu^{-\alpha}$ and $\alpha = 1$. Along the polar direction, a bi-conical, ionized wind with 
a 25$^\circ$ half-opening angle\footnote{The half-opening angle of our model is smaller than what is observed \citep{Goosmann2011}.
We will change this value when the comparison between MontAGN and STOKES will be completed, see Paper~I.} with respect to the polar 
axis flows from the central source to 25~pc. The wind is assumed to be ionized and therefore filled with electrons. It is optically 
thin (optical depth in the V-band along polar direction $\tau_{\rm V}$ = 0.1). Along the equatorial plane, a flared disk sets on at 
0.05~pc (a typical dust sublimation radius, see \citealt{Kishimoto2007}) and ends at 10~pc. The half-opening angle of the dust structure 
is fixed to 30$^\circ$ \citep{Marin2012} and its V-band optical depth is of the order of 50. The dust is composed of 100~\% silicates 
with grain radii ranging from 0.005~$\mu$m to 0.25~$\mu$m, together with a size distribution proportional to $a^{\rm -3.5}$ ($a$ being 
the grain radius).

The observer's viewing angle is set to 90$^\circ$ with respect to the symmetry axis of the model. More than 10$^7$ photons were sampled
to obtain polarimetric images with a pixel resolution of 1~pc (9 milli-arcsec at 14.4~Mpc). For this proceedings note, we selected the 
images computed at 1~$\mu$m and used the Monte Carlo code STOKES.

\subsection{Results of our baseline model}

\begin{figure}[ht!]
 \centering
 \includegraphics[trim = 0mm 0mm 0mm 0mm, clip, width=8.3cm]{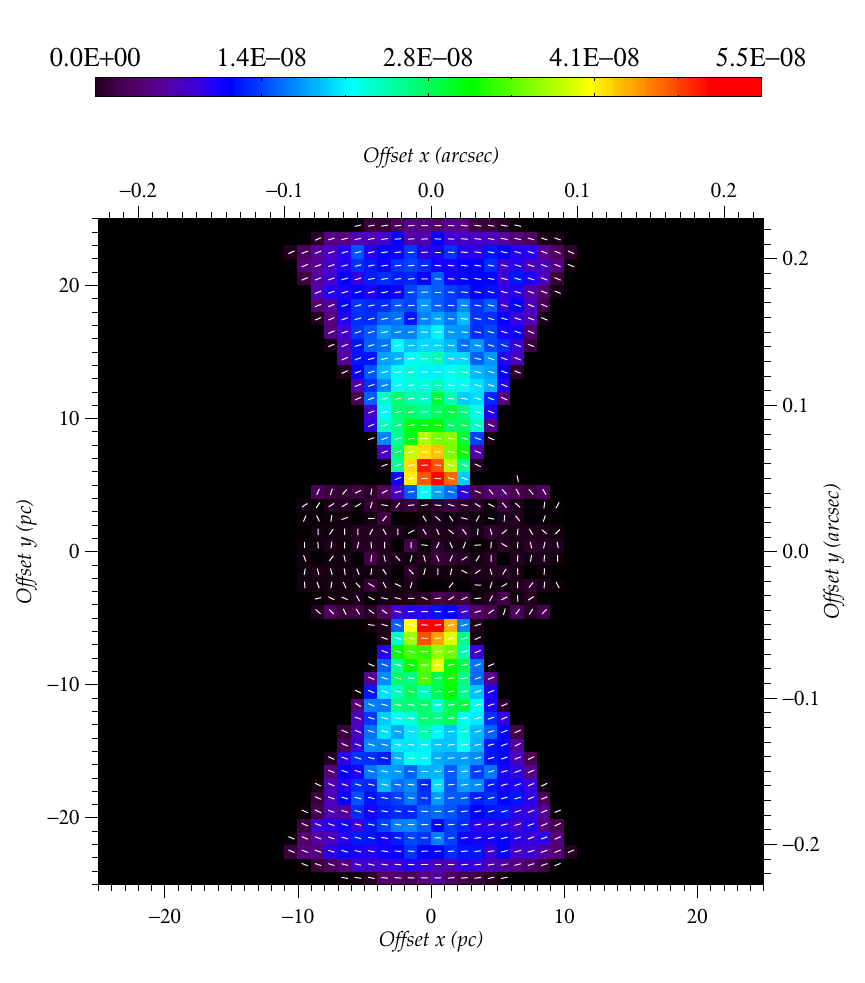}      
 \includegraphics[trim = 0mm 0mm 0mm 0mm, clip, width=8.3cm]{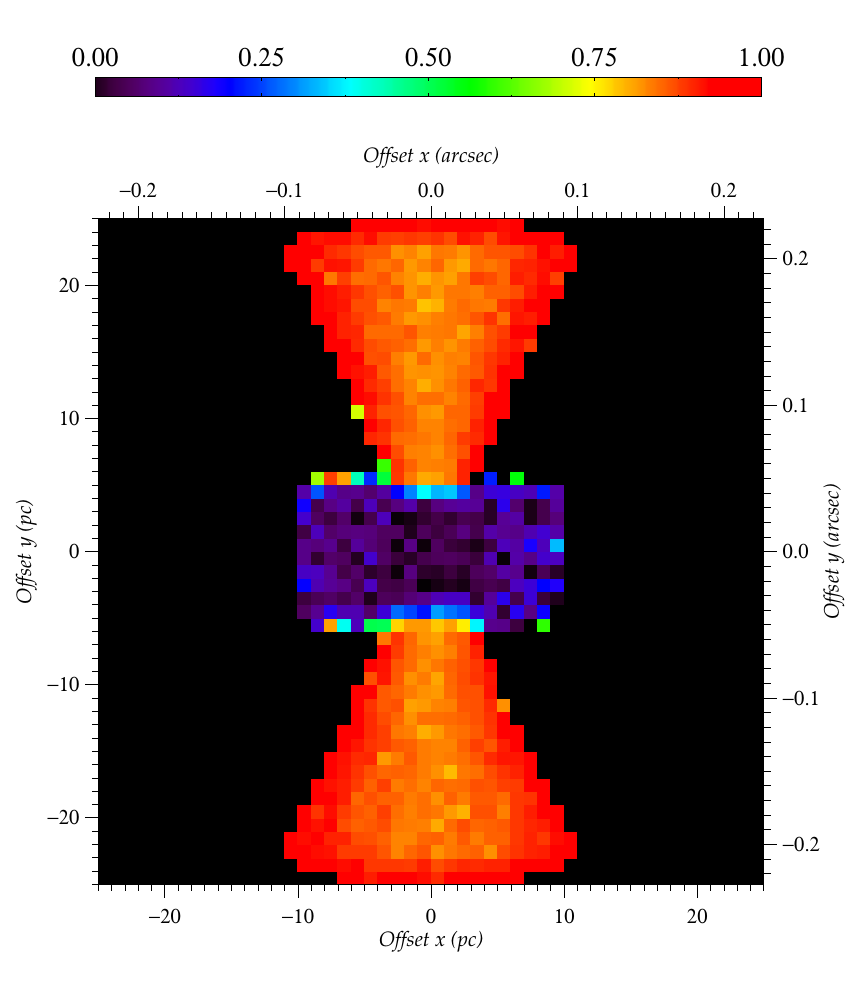} 
  \caption{1~$\mu$m polarimetric simulations of NGC~1068. Left figure 
	    shows the color-coded polarized flux in arbitrary units
	    with the polarization position angle superimposed
	    to the image. Right figure shows the color-coded degree 
	    of polarization (from 0, unpolarized, to 1, fully polarized).}
  \label{marin:fig1}
\end{figure}

The polarimetric maps for our baseline model are shown in Fig.~\ref{marin:fig1}. The left image shows the polarization position angle 
superimposed to the 1~$\mu$m polarized flux. The polarization vectors show the orientation of polarization but are not proportional to 
the polarization degree (which is shown in the right image). 

The polar outflows, where electron scattering occurs at a perpendicular angle, shows the strongest polarization degree (up to 90 -- 100~\%)
associated with a centro-symmetric polarization angle pattern. This in perfect agreement with the polarization maps taken by the optical and 
infrared polarimeters in the 90s and recently upgraded by \citet{Gratadour2015}. At the center of the model, the photon flux is heavily 
suppressed by the optically thick material, but scattered radiation from the cones to the surface of the torus leads to a marginal flux 
associated with a weak polarization degree\footnote{Polarization degrees quoted in the text are for the scattered-induced polarization only.
Dilution by the interstellar polarization, host galaxy starlight and starburst light will drastically reduce the observed polarization degree.}
($<$~8~\%). Such degree of polarization is in very good agreement with the values observed by \citet{Gratadour2015} at the location of the 
nucleus (5 -- 7~\%). However, compared to the results of the previous authors, the polarization position angle from the modeling is almost 
centro-symmetric rather than aligned/anti-aligned with the circumnuclear dusty structure. It is only on the highest point of the torus morphology 
that a $\sim$~50~\% polarization degree associated with a higher flux can be found, due to a lesser amount of dust facing the wind-scattered 
photon trajectories. When the whole picture is integrated, the resulting polarization degree is of the order of 60~\%.

\subsection{Accounting for the ISM}

\begin{figure}[ht!]
 \centering
 \includegraphics[trim = 0mm 0mm 0mm 0mm, clip, width=8.3cm]{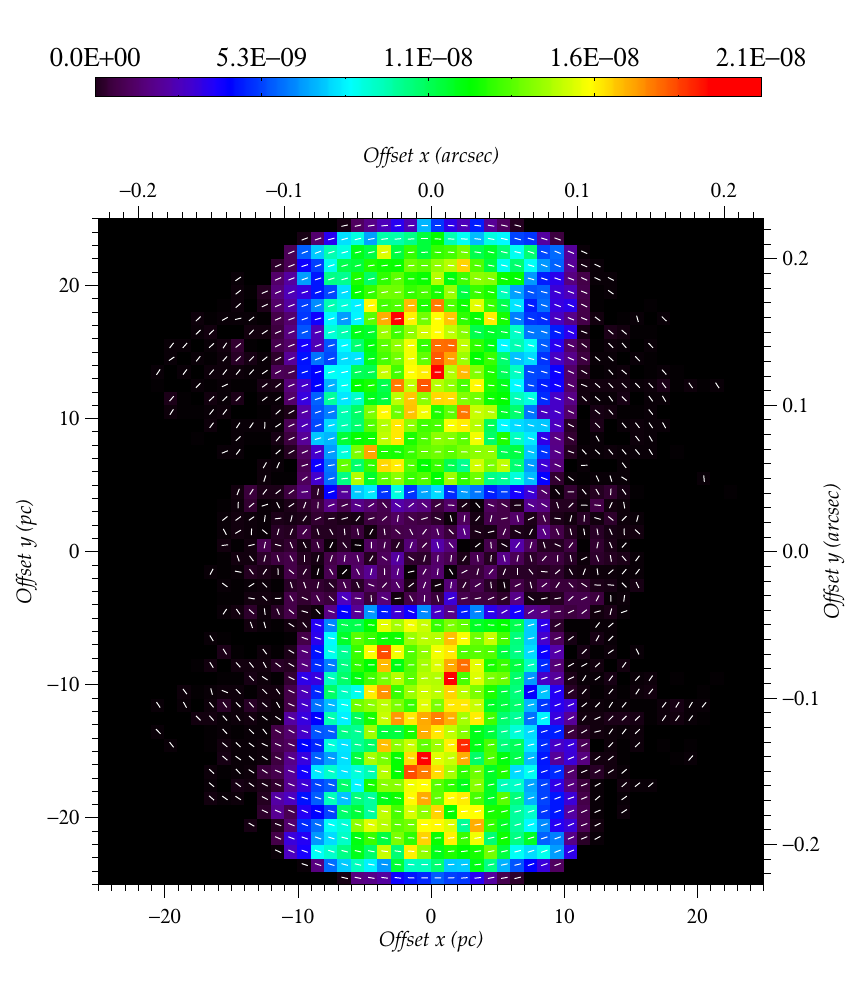}      
 \includegraphics[trim = 0mm 0mm 0mm 0mm, clip, width=8.3cm]{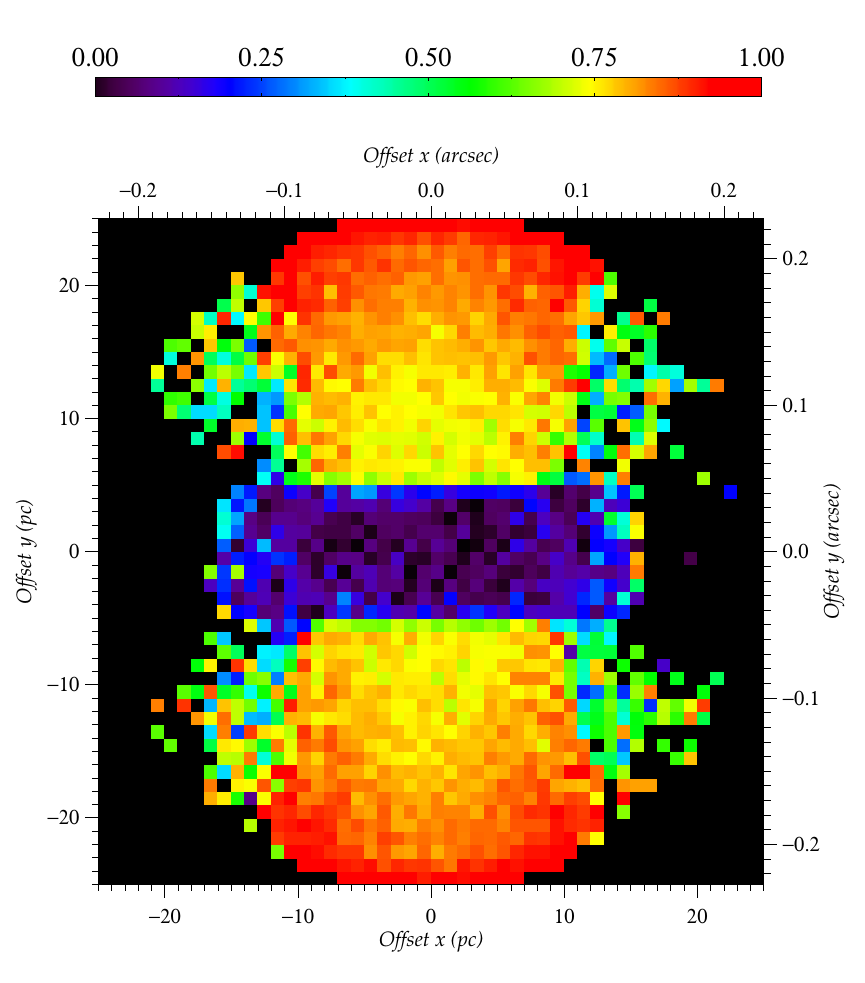} 
  \caption{Same as Fig.~\ref{marin:fig1} with the addition of 
	    optically thin interstellar matter around the model}
  \label{marin:fig2}
\end{figure}

To test multiple configurations, we ran a second series.Based on the previous setup, we included interstellar matter (ISM) around 
the model. The ISM dust grains share the same composition as the dust in the circumnuclear AGN region and the ISM is optically thin 
in all directions ($\tau_{\rm V}$ = 0.5). Our new polarimetric images are shown in Fig.~\ref{marin:fig2}

A striking difference between the ISM-free (Fig.~\ref{marin:fig1}) and the ISM-included (Fig.~\ref{marin:fig2}) polarimetric maps
is the shape of the outflowing winds seen in transmission through the dust. The perfect hourglass shape observed when the AGN was in 
vacuum is now disturbed. The global morphology is more similar to a cylinder, with no flux gradient observed as the photons propagate
from the central engine to the far edges of the winds. The overall flux distribution is more uniform throughout the winds, which seems
to be in better agreement with what has been observed in the same band, at least for sub-arcsec scales \citep{Packham1997}. At better
angular resolutions, polarized flux images are still needed. The polarization position angle has retained its centro-symmetric pattern
but the polarization angle is more chaotic at the location of the torus, due to additional dust-scattering. The degree of polarization 
at the center of the AGN is the same as previously but the integrated map shows a slightly smaller polarization degree (58~\%) due to 
depolarization by multiple scattering.

\section{Discussion}
Running our radiative transfer codes at 1~$\mu$m, we found that a NGC~1068-like model produces the centro-symmetric polarization 
angle pattern already observed in the optical and infrared bands. Disregarding additional dilution by other sources, the polarization
position angle pinpoints the source of emission. Including the ISM in the model does not change the results but tends to decrease 
the final polarization degree. It also changes the flux repartition in the outflowing winds that act like astrophysical mirrors, 
scattering radiation from the hidden nucleus. Compared to what has been shown in \citet{Gratadour2015} in the H (1.6~$\mu$m) 
and K (2.2~$\mu$m) bands, we find very similar levels of linear polarization at the center of the model, where the central engine is 
heavily obscured by dust. However, we do not retrieve the distinctive polarization position angle of the torus found by the authors. 
According to our models, the pattern is at best centro-symmetric rather than directed perpendicular to the outflowing wind axis. 

Additional work is needed to explore how such a distinctive pattern can arise at the location of the torus. In particular, adopting the 
most up-to-date morphological and composition constraints from literature is mandatory to built a coherent NGC~1068 model. Including
effects such as polarization by absorption (dichroism) will be necessary. Comparing our results with past linear and circular polarization 
measurements (e.g, \citealt{Nikulin1971,Gehrels1972,Angel1976}) will drive our models towards the right direction. Finally, the broadband 
coverage of the codes, from the X-rays to the far infrared, will allow us to robustly test our final model against spectroscopic and 
polarimetric observations in many wavebands.

\begin{acknowledgements}
The authors would like to acknowledge financial support from the Programme National Hautes Energies (PNHE).
\end{acknowledgements}

\bibliographystyle{aa}  
\bibliography{marin} 

\end{document}